\documentclass{article}
\usepackage{ijcai26}

\usepackage{times}
\usepackage{soul}
\usepackage{url}
\usepackage[hidelinks]{hyperref}
\usepackage[utf8]{inputenc}
\usepackage[small]{caption}
\usepackage{graphicx}
\usepackage{amsmath}
\usepackage{amsthm}
\usepackage{booktabs}
\usepackage{algorithm}
\usepackage{algorithmic}
\usepackage{color}
\usepackage{multirow}
\usepackage{enumitem}
\usepackage[switch]{lineno}

\title{Fine-grained Classification of A Million Life Trajectories from Wikipedia}

\author{
Zhaoyang Liu\and
Xiaocong Du\and
Yixi Zhou\and
Ye Shi\And
Haipeng Zhang\\
\affiliations
ShanghaiTech University\\
\emails
\{liuzhy2023, duxc2023, zhouyx2022, shiye, zhanghp\}@shanghaitech.edu.cn
}

\begin{document}

\maketitle
\begin{abstract}
Life trajectories of notable people convey essential messages for human dynamics research. These trajectories consist of (\textit{person, time, location, activity type}) tuples recording when and where a person was born, went to school, started a job, or fought in a war. However, current studies only cover limited activity types such as births and deaths, lacking large-scale fine-grained trajectories. Using a tool that extracts (\textit{person, time, location}) triples from Wikipedia, we formulate the problem of classifying these triples into 24 carefully-defined types using textual context as complementary information. The challenge is that triple entities are often scattered in noisy contexts. We use syntactic graphs to bring triple entities and relevant information closer, fusing them with text embeddings to classify life trajectory activities. Since Wikipedia text quality varies, we use LLMs to refine the text for more standardized syntactic graphs. Our framework achieves 84.5\% accuracy, surpassing baselines. We construct the largest fine-grained life trajectory dataset with 3.8 million labeled activities for 589,193 individuals spanning 3 centuries. In the end, we showcase how these trajectories can support grand narratives of human dynamics across time and space. Code/data are publicly available.
\end{abstract}

\section{Introduction}

Life trajectories~\cite{elder1994time} of notable individuals capture their activities across time and space, offering valuable insights into human dynamics. These trajectories often coincide with the rise of new technologies, the spread of ideas, and the formation of innovative hubs~\cite{schatzki2019social,schatzki2022trajectories,verginer2020cities,atari2025chronospatial}. A pioneering study in \textit{Science}, such as an analysis of birth and death locations of over 300,000 notable figures, have uncovered long-term patterns in the emergence and evolution of cultural centers. Research also focuses on specific types and activities—such as scientists’ career movements, mobility of creative talents, labor flows in tech industries, and the travels of artists and politicians—to understand related social, economic, and political developments~\cite{huang2020careerhistorical,dickinson2015identifying,herrera2023ballet,park2019global,creativityovertime}. With advances in LLM agents, such data now supports simulating historical figures and interactively exploring their lives and contributions~\cite{bai2024baijia}.

However, previous data-driven studies are limited to certain types of activities, such as births and deaths~\cite{schich2014network}, hi-tech jobs~\cite{park2019global}, and scientific careers~\cite{huang2020careerhistorical}, which are relatively easier to obtain from sources including Freebase, LinkedIn and scientific papers. Data of various activity types, such as political visits~\cite{aleksanyan2021state}, general careers~\cite{wang2019early}, marriages~\cite{wolfinger2008problems}, education~\cite{quarles2020shape}, and competitions~\cite{mukherjee2019prior}, albeit important, is scarce, often hand-collected. 

Though sources such as Wikipedia's biography pages contain rich information for extracting $(person, time, location)$ triples that represent trajectories, simply applying Named Entity Recognition (NER) tools and combining the candidate entities often results in inaccuracies, with a precision of around $30\%$~\cite{zhang2024paths}. Consequently, researchers specifically design COSMOS, a system that performs extraction with high F1 scores, outputting the triples as well as the context sentences that contain the triples~\cite{zhang2024paths}.
This mitigates the data scale problem and serves as a nice groundwork for our study. But still, obtaining fine-grained trajectory activity types remains an unsolved yet challenging problem.
We decide that it should not be formulated into an unsupervised learning problem, since clustering methods for these tasks~\cite{zhang2021model,qian2023open} are likely to give uninterpretable results when there is a large variety of activity types. 

\begin{figure}
    \centering
    \includegraphics[width=\linewidth]{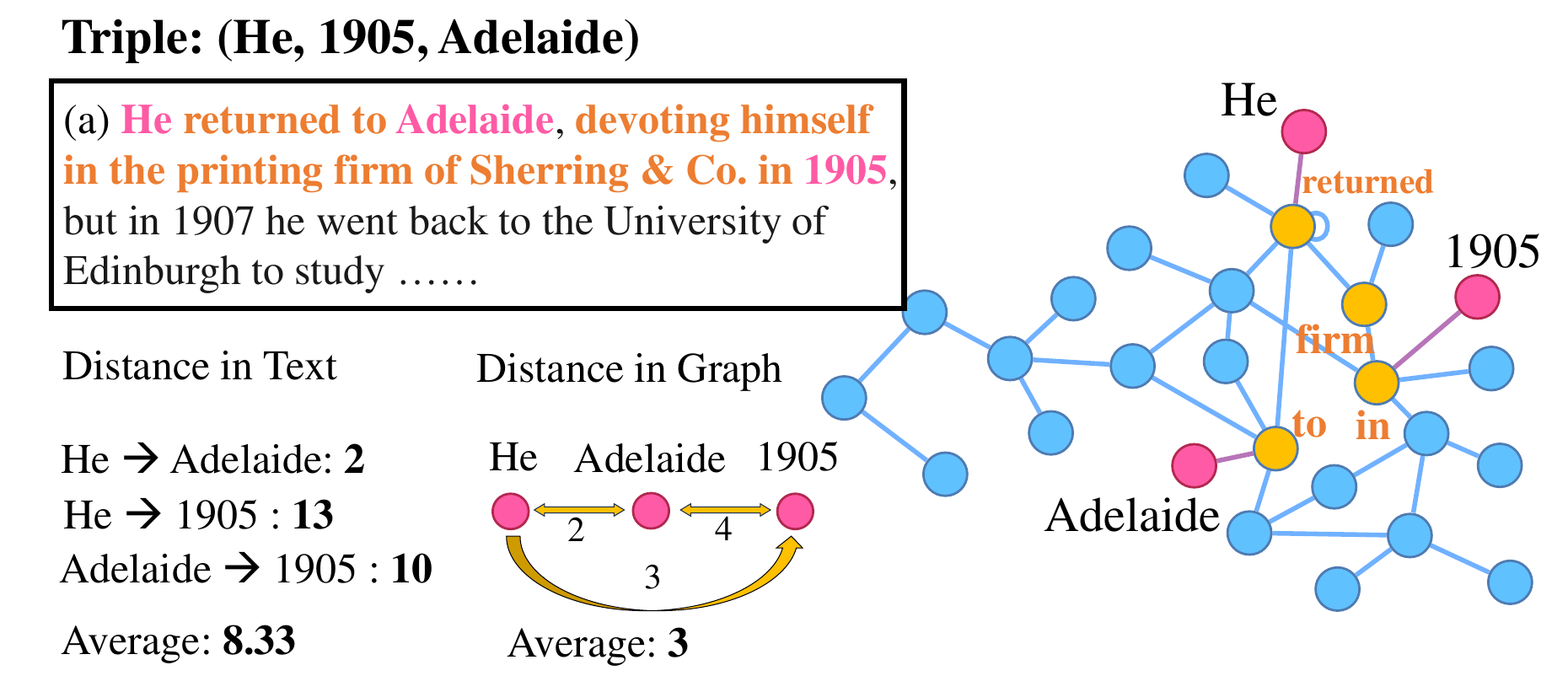}
    \caption{Examples of life trajectories extracted from Wikipedia. Words in pink are the extracted entities in triples indicating \textbf{person, time and location}. The graph is the syntactic graph constructed by SpaCy. The red nodes correspond to extracted entities in triples. The orange ones are nodes and words on the paths.}
    \label{fig:examples}
\end{figure}

\textbf{We formulate it into a multi-classification problem -- given a triple of $(person, time, location)$ and its context sentence from Wikipedia (i.e., the output of the COSMOS extraction tool~\cite{zhang2024paths}), decide the type of its trajectory activity.} 
For example, as shown in Figure~\ref{fig:examples}, given the triple (`He', `1905', `Adelaide') and its corresponding context sentence (a), our objective is to classify the type of this specific life trajectory activity (career).
As a necessary pre-step, we draw upon the types of ACE~\cite{doddington2004automatic} event extraction dataset and related work of life trajectory analysis~\cite{herrera2023ballet,huang2020careerhistorical,yen2019personal-bak}, and propose a taxonomy including \textbf{24 types of life trajectory activities in 9 categories}, shown in Table~\ref{tab1}.

This task differs from regular text classification~\cite{minaee2021deep} in a sense that instead of classifying mere text, we are actually classifying triples, with the sentences they belong to as the context. This requires reference to both sentence and triple. The triple should be well emphasized and the semantic relations between the triple entities and their related text should be aggregated. 
Take the triple and its context sentence in Figure~\ref{fig:examples} as an example.
Without the triple, the sentence does not indicate a clear activity type. The first half of the sentence suggests the career type while the second half is about education. Only by situating the triple into (the first half of) the sentence can we locate the really related information of the trajectory types and confidently identify it as a career activity. Such a constraint is similar to the task of aspect-level sentiment classification (ALSC)~\cite{schouten2015survey,fei2022inheriting} which is to classify the sentiment associated with a given entity among multiple entities in the text. For ALSC, researchers utilize a syntactic graph to capture the association between sentimental words and the given entity, and model their distance on the graph~\cite{dai2021does,zhang2020convolution}. This inspires us to use a syntactic graph, but to bring text information relevant to the triple entities closer, for better emphasis and aggregation. As we can see in Figure~\ref{fig:examples}, the three triple entities (`He', `1905', `Adelaide') are scattered in the sentence (the average pairwise distance is 8.33 words), leaving plenty of space for noisy words to appear in between. Within the syntactic graph, the average pairwise distance between the triple entities is largely shortened (to 3). As a result, we can pay more attention to the words on the shortest paths between the triple entities with reduced noise. 
ALSC studies, however, simply concatenate train text representation and syntactic structure embedding without fusing them during the whole training process, which leads to inefficient utilization.

Attention mechanism~\cite{vaswani2017attention} aims to focus on the most relevant information in the entire sequence and its MASK~\cite{devlin2018bert} helps focus on the important words by masking words that do not matter. This inspires us to transform the syntactic structure of a sentence into a MASK vector that guides language models to associate textual pieces according to how they are structurally related and output text embedding fused with syntactic structure. As mentioned previously, the words in the shortest paths among the triple entities on the syntactic graph are the ones we want to emphasize, forming a local \textbf{subgraph} of the overall syntactic \textbf{graph}. This subgraph's corresponding locations are `unmasked' in MASK, guiding the text embedding to focus on the triple and its relevant information. Though these embeddings represent the words connected to the triple entities, other words in the sentence are not included. We therefore combine the regular textual embedding to obtain a more comprehensive representation of a life trajectory activity.

We note that Wikipedia, being collaboratively edited, lacks a consistent writing style, as observed in other studies~\cite{liu2021can}. This results in significant variation in syntactic structures, even for sentences of the same type, introducing potential noise. To address this, we use LLMs as a neat tool to rewrite sentences in a more standardized manner, enabling more uniform syntactic graphs and reducing noise caused by diverse writing styles.

In this paper, we propose the SAM4LTC (\textbf{S}yntax \textbf{A}ware \textbf{M}asked model for \textbf{L}ife \textbf{T}rajectory \textbf{C}lassification) framework to fuse syntactic information into text representations for fine-grained classification of life trajectory activities. Its input is the life trajectory triple and its context sentence 
{extracted from Wikipedia by}
COSMOS~\cite{zhang2024paths} and the syntactic structure of the sentence is refined using LLM. It first constructs the syntactic graph of the life trajectory activity sentence and then extracts the subgraph of the words most related to the triple by searching the shortest paths among the triple elements. Finally, additional supervised contrastive learning loss is adopted to enhance the model's ability to learn to distinguish different classes. To train SAM4LTC, we manually annotate 2,826 life trajectory activities randomly selected from Wikipedia bio pages. We summarize the contributions of our work as follows:
\begin{itemize}
\item We propose a new life trajectory labeling task and a taxonomy of life trajectory activities, as well as a manually annotated dataset on this taxonomy. 

\item 
The SAM4LTC framework fuses syntactic and semantic information achieving an Accuracy of 84.5\%. Additionally, we refine syntactic structures with LLM to improve data quality.
\item We publicly release our model, manually-annotated dataset, and the largest fine-grained life trajectory dataset to date (3.8 million tuples)\footnote{\url{https://anonymous.4open.science/r/SAM4LTC-B7C9/} We release 247,292 tuples in this version; all 3.8M tuples will be released upon acceptance.} to support research on human dynamics and information extraction. Analysis of 589,193 individuals across 3 centuries demonstrates how fine-grained trajectories enable new insights into human dynamics.

\end{itemize}

\section{Related Work}

\subsection{Life Trajectories and Human Dynamics}
Analysis of life trajectories brings insights into various fields regarding human dynamics. For example, births and deaths of more than 300,000 people in history are analyzed and the cultural evolutionary process is outlined based on their life migration~\cite{schich2014network}. Similar analysis~\cite{creativityovertime} of artists also reveals the close connection between the immigrants in the region and the rise and fall of creativity centers. 

Advances of interactive human and large-scale social simulation based on LLM agents, life trajectory data now holds value beyond analytics~\cite{shanahan2023role,park2023generative}. Researchers have compiled profiles of historical figures—including details such as birth, death, and official posts with corresponding times and locations—which are used to fine-tune LLMs for accurate role-playing and interactive exploration of their lives and contributions~\cite{bai2024baijia}.

Efforts on extracting life trajectories on a large scale are made to alleviate the scarcity of such data and provide a substantial data source for related analysis to explore life trajectories more comprehensively~\cite{zhang2024paths}. Their work helps us to extract high-quality life trajectory triples from Wikipedia using their well-developed tools. Multiple life trajectory activities are needed to construct a complete life course of a human.

\subsection{Sentence Classification using Syntactic Graph}
Traditional sentence classification utilizes pre-trained language models~\cite{devlin2018bert,yang2019xlnet} (PLMs) to learn the overall sentence embeddings, achieving good performance with the external knowledge learned during pre-training. Meanwhile, there are also research modeling sentences into graphs to mine the internal structure of the sentences. For example, TextGCN~\cite{yao2019graph} constructs the graph of a sentence based on its word co-occurrence and sentence-word relations, capturing higher order neighbors' information. GatedGCN~\cite{lai2020event} uses the syntactic graph to model a sentence and apply Graph Convolutional Network (GCN) with a gating mechanism controlling the information flow to learn the embeddings most relevant to the target words. 

In our scenario, we observe that triple entities are closer to each other in the syntactic graph while far away in a sentence. Therefore, we use the syntactic graph to guide the model to focus on the life trajectory triples in the context sentence.

\subsection{Aspect-Level Sentiment Classification}

Studies of ALSC focus on identifying the associated sentiment of a specific entity in texts rather than taking the texts as a whole. To do so, they usually go beyond sentence embeddings and further explore the syntactic graph~\cite{ke2021incorporating,huang2020syntax}. For instance, paths between the aspect word and sentimental word on the syntactic graph are searched and transformed into path sequence which is learned to get the structure by a sequence encoder~\cite{ke2021incorporating}.
GAT is used to learn the syntactic information~\cite{huang2020syntax} with the fine-tuned word embeddings of PLMs as their node features~\cite{wang2020relational}.

In our task, we agree that both text embeddings and syntactic structures are useful in classifying life trajectory activities. However, usage of syntactic structures in the above work is inefficient for not fusing the structure information and text embeddings during the whole training process. To fuse them deeper, we introduce a MASK generation method based on syntactic graph making our model focus on life trajectory triple among multiple entities. 

\section{Task Definition}
Given a Wikipedia bio page, we use COSMOS to extract the sentence $S$, and the triple $T = (person, time, location)$, where the elements in the triple $T$ are in the sentence $S$, the sentence describes the life trajectory activity of the $person$ with additional information about $time$ and $location$. Specifically, the $person$ is experiencing his/her life event in the $location$ at the $time$. In addition, the label space $ L = (l_1, l_2, \cdots,l_n)$ represents the taxonomy of the life trajectory activities shown in Table~\ref{tab1}. We train a model $f$ with trainable parameter $\Theta$ to assign a label (i.e., the type of the life trajectory activity that feeds into $f$) in $l \in L$ to each triple $T$ based on the sentence $S$:

\begin{equation}
    l = f(S, T, \Theta)\label{eq}
\end{equation}

The label space is introduced in the Dataset section.


\section{Dataset}


\subsection{Data Collection}

Our data is from the biography pages of English Wikipedia\footnote{The Wikipedia dumps from \url{https://dumps.wikimedia.org/}.}. We first randomly select candidate pages for extraction using the entities with property \textit{instance of human} in Wikidata.

In all, we randomly select 2,000 biography pages for extraction. After obtaining the candidate biographical pages, we use the COSMOS tool~\cite{zhang2024paths} to extract life trajectory triples from those selected pages and keep the sentence that each triple is from. 

\subsection{Taxonomy and Annotation of Life Trajectories}
As mentioned in introduction, existing research mainly analyzes births and deaths, and has not yet specifically designed a taxonomy for life trajectory activities. Therefore, drawing on the types of events in the task of event extraction ACE~\cite{doddington2004automatic}, studies in related fields~\cite{herrera2023ballet,huang2020careerhistorical,janssen2011age}, and our observation of the data, we propose the following taxonomy of life trajectory activities\footnote{See section A.1 of Supplementary Materials for details of the types and annotation.} consisting of 24 types in 9 categories. We note that this taxonomy may be to some extent arbitrary, and further research may need to adjust it according to specific needs.

\begin{table}[htbp]
\caption{Taxonomy of Life Trajectory Activities} 
\begin{center}
\begin{tabular}{p{1.3cm}p{5.5cm}ll}
        \toprule
        Category & Type\\
        \hline
        Life & Birth, Death, Marriage, Injure and Illness, Settlement, Education, Give birth, Accident, Purchase and Sell, Divorce\\
        \hline
        Career & Career\\
        \hline
        Pro-Event & Competition, Performance, Exhibition, Creation, Campaign, Start org\\
        \hline
        Contact & Meet, Assembly\\
        \hline
        Justice & Justice\\
        \hline
        Attack & Attack\\
        \hline
        Movement & Movement\\
        \hline
        Military & Military\\
        \hline
        Other & Other\\
        \bottomrule
\end{tabular}
\label{tab1}
\end{center}
\end{table}

\begin{figure}
    \centering
    \includegraphics[width=\linewidth]{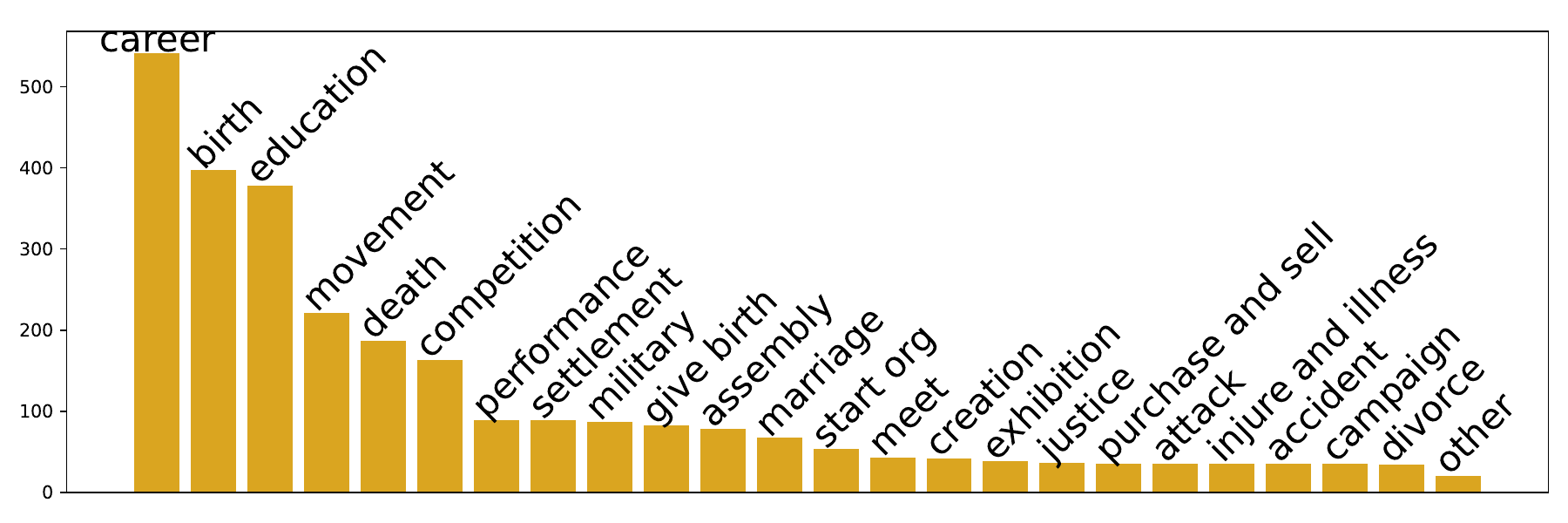}
    \caption{Distribution of types in the \textit{Regular} dataset.}
    \label{fig:dis}
\end{figure}

Based on the taxonomy, we randomly sample and manually annotate a dataset of 2,826 samples from the extracted life trajectory triples for training and testing our model. In addition, the distribution of our dataset is shown in Figure~\ref{fig:dis}. In the annotated dataset, the most frequent life trajectory activity is related to career, followed by birth and education. 

\subsection{Syntax Refinement using LLM}

Previous studies use LLM to annotate their data~\cite{whitehouse2023llm,moller2024parrot} or directly generate data samples satisfying their requirements~\cite{li2023synthetic,ho2022large,xu2025largereasoningmodelssurvey} to augment their training data for better performance of their models.
Unlike these studies, we do not seek to generate new data, but to refine the syntactic structures of the sentences by requiring LLM to rewrite the sentences in another way without changing the original meaning. This is because we are concerned that training on samples generated by LLM may lead to overfitting issues, making it difficult to generalize on real-world samples. Simply rewriting the sentences preserves real-world data distribution while providing a more standardized syntactic structure for our proposed model to learn.
Specifically, we require the LLM to rewrite the sentence $S$ to get the refined sentence $S_{ref}$:

\begin{equation}
S_{ref} = LLM(S, prompt)
\end{equation}

The prompt (Few-Shot) can be found in Supplementary Materials (Section D.3). 
We refer to the fully manually annotated dataset as the \textit{regular} dataset, while the dataset refined by the LLM is called the \textit{LLM-Refined} dataset.
Notably, when
GPT-4 occasionally fails to rewrite a sentence, we implement a simple quality control mechanism by prompting it to redo the task until the sentence has been successfully modified.

\section{Method}

\begin{figure*}
    \centering
    \includegraphics[width=\linewidth]{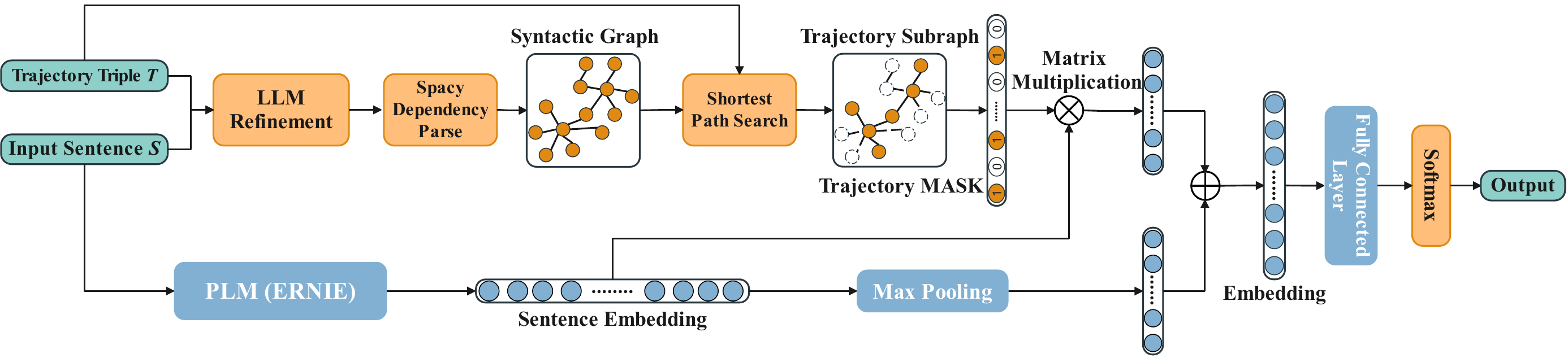}
    \caption{Workflow of SAM4LTC. The trainable modules and embeddings are in blue.}
    \label{fig:SAM-ERNIE}
\end{figure*}

\subsection{Overview of the Method}

The workflow of SAM4LTC is shown in Figure~\ref{fig:SAM-ERNIE}. Given a sentence $S$ and a triple $T$, we first use the LLM to refine the sentence without modifying the words in the triple and changing the meaning of the sentence and add special tokens for the elements in the triple to make the language model understand the elements in the triple as a whole. Then we construct a graph and a trajectory \textbf{subgraph} using $S$ and $T$ with nodes representing tokens tokenized by a PLM and edges representing the dependency relations parsed by SpaCy\footnote{\url{https://spacy.io/}}, a language processing tool for constituency and dependency parsing. 

Based on the trajectory \textbf{subgraph}, we generate a `MASK' vector to get the embedding of the words most relevant to the trajectory activity and concatenate it with the sentence embedding output by ERNIE. Finally, as a standard procedure, a fully connected layer is used to get the classification results. 

One preferred option for the aforementioned PLM is ERNIE~\cite{sun2019ernie}. As a high-ranked and open-sourced PLM on the GLUE~\cite{wang2018glue} leaderboard, it well emphasizes entity information during its pre-training process, and this would help better understand the person, time and location entities in the triple for our scenario.

\subsection{Graph Construction}
For the input sentence $S = \{w^1, w^2, ..., w^m\}$ with length $m$ and triple $T = \{ t_p, t_t, t_l \}$ extracted from $S$, we first cut them into subwords with ERNIE tokenizer, which become:
\begin{align}
    S &= \{w^1, w^{1\#1}, \cdots, w^m, \cdots, w^{m\#k}\}, \\
    T &= \{t_p^1, \cdots, t_p^{p\#k}, t_t^1, \cdots, t_t^{t\#k}, t_l^{1}, \cdots, t_l^{l\#k}\},
\end{align}
where $w^{m\#k}$ is the $k$-th subword of $w^m$ and $t_p$, $t_t$, $t_l$ is the words indicating person, time and location with word length of $p$, $t$, $l$. Then we add special tokens to the vocabulary of the tokenizer to mark the corresponding triple elements in the sentence $S$.

Upon tokenizing the given sentence and words in triple into subwords, we utilize SpaCy to construct a graph $G = (V, E)$ of the sentence based on the syntactic structure and dependency relations within the sentence parsed by SpaCy. In this overall graph, nodes correspond to the subwords or tokens in $S$, and edges denote the dependencies between subwords. Additionally, we add another type of edge representing words segmented by the ERNIE tokenizer, which is:

\begin{gather}
V = S, \\
B = \{ (w^1, w^{1\#1}), (w^1, w^{1\#2}) \cdots,  (w^n, w^{m\#k})\}\\
E = P \cup B
\end{gather}
where $P$ is the result of dependency parsing by SpaCy.

Verbs typically serve as the bridge to connect the entities in a sentence~\cite{viberg1983verbs}. Hence, we search the shortest paths between the words in $T$ and their nearest verbs and use the words on the paths to extract a local subgraph from the overall graph to represent the connection among the trajectory triple. Specifically, let $SP$ be the shortest path algorithm and $Verb$ be the verbs in the sentence identified by SpaCy, then compute 3 paths:

\begin{align}
&Path(t) = min(SP(t, Verbs)), \\
&Path = Path(t_p) \cup Path(t_t) \cup Path(t_l),
\end{align}
where $SP(A, B)$ returns all the shortest paths between nodes in $A$ and nodes in $B$, $t_p$, $t_t$, $t_l$ are the nodes corresponding to person, time and location in $T$. 

Then the local subgraph $G^{\prime} = (V^{\prime}, E^{\prime})$ can be extracted from the overall graph using the nodes in $Path_p$, $Path_t$ and $Path_l$:

\begin{align}
&V^{\prime} = Path, \\
&E^{\prime} = \{(u, v)\}, \, u,v \in V^{\prime},
\end{align}

Note that each sentence extracted from Wikipedia consist limited words and therefore the additionally introduced computation is trivial comparing to the neural network though the $O(n^3)$ shortest path algorithm is adopted in the framework.

\subsection{Representation with Generated MASK }

After obtaining the local subgraph, we compute the `MASK' vector by simply marking the corresponding position as $1$ if the corresponding subword is in $G$ or $V^{\prime}$ and $0$ otherwise:

\begin{equation}
MASK[i] = 
  \begin{cases}
    1 & \text{if } i\text{-th token of }S \in V^{\prime}, \\
    0 & \text{otherwise,}
  \end{cases}
\end{equation}

\begin{table*}[htbp]

    \caption{Performance comparisons for life trajectory activity classifications. The best ones are in bold.} 
    \begin{center}
    \begin{tabular}{c|ccccccccc}
    \toprule
            \multirow{2}{*}{Model} & \multicolumn{3}{c}{Regular/Type} & \multicolumn{3}{c}{LLM-Refined/Type} & \multicolumn{3}{c}{LLM-Refined/Category}\\
            \cmidrule(lr){2-4} \cmidrule(lr){5-7} \cmidrule(lr){8-10}
                & P & R (Acc) & F1 & P & R (Acc) & F1 & P & R (Acc) & F1 \\
            \hline
            Bi-LSTM & 55.7\% & 61.0\% & 57.5\% & 61.6\% & 60.6\% & 60.4\%  & 66.1\% & 66.7\% & 66.4\%\\
            TextGCN & 62.1\% & 61.3\% & 29.3\% & 63.4\% & 61.5\% & 60.3\%  & 76.8\% & 71.7\% & 71.7\%\\
            R-GAT & 74.1\% & 78.0\% & 74.8\% & 71.3\% & 75.8\% & 72.2\% & 81.4\% & 81.2\% & 80.7\%  \\
            BERT & 82.5\% & 81.6\% & 81.4\% & 83.9\% & 83.0\% & 82.8\% &  84.7\% & 84.7\% & 84.6\%\\
            XLNet & 82.5\% & 81.7\% & 81.4\%  & 84.2\% & 83.3\% & 83.0\% & 85.6\% & 85.8\% & 85.7\%\\
            ERNIE & 83.1\% & 82.1\% & 81.8\% & 84.2\% & 83.2\% & 82.9\% &  85.7\% & 85.9\% & 85.8\%\\
            GPT-4 & 73.9\% & 76.2\% & 72.6\% & 75.1\% & 72.6\% & 70.0\%  & 80.6\% & 81.5\% & 80.5\%\\
            GPT-5 & 71.2\% & 73.5\% & 70.0\% & 74.7\% & 74.6\% & 72.9\%  & 78.8\% & 79.6\% & 78.4\%\\
            EvoPrompt & 74.2\% & 71.6\% & 70.8\% & 76.1\% & 73.9\% & 74.3\%  & 80.6\% & 81.5\% & 80.5\%\\
            SAM4LTC & \textbf{83.7\%} & \textbf{82.5\%} & \textbf{82.2\%} & \textbf{85.2\%} & \textbf{84.5\%} & \textbf{84.5\%} &  
            \textbf{86.8\%}& \textbf{87.0\%}& \textbf{86.9\%}\\
\bottomrule
    \end{tabular}
    \label{tab1:perf}
    \end{center}
\end{table*}

To get the representation of the trajectory activity, we perform matrix multiplication on the `MASK' vector and the embeddings output by ERNIE $h_e$. The trajectory activity embedding $h_1$ is obtained by masking the $MASK$ on $h_e$:

\begin{align}
h_1 = MM(h_e, MASK), \\
h_1^\prime = MeanPooling(h_1),
\end{align}
where $MM$ is the matrix multiplication, and $MeanPooling$ is the mean-pooling operation. 

The contextual embeddings are obtained the through max-pooling of $h_e$:

\begin{align}
h_2 = MaxPooling(h_e)
\end{align}
where $MaxPooling$ is max-pooling operation.

Finally, the embedding used for contrastive learning and cross entropy loss is obtained by concatenating $h_1$ and $h_2$:
\begin{align}
h_{scl} = h_1 \oplus h_2,
\end{align}

\subsection{Contrastive Learning and Cross-Entropy Loss}

Conventional cross-entropy loss is inadequate in our scenario of insufficient annotated data~\cite{zhang2018generalized}, for the possibility of overfitting and poor generalization. Supervised contrastive learning (SCL) integrates the idea of ``learn to compare" into supervised learning~\cite{khosla2020supervised}, mitigating the issue by reducing reliance on labels and improving the model's generalization ability.

For the output feature vector $h_{scl}$, we compute the supervised contrastive learning loss as follow:
\begin{equation}
    L_{scl} = \sum_{i=1}^N \frac{-1}{N_{y_i} - 1} \sum_{j=1}^N \textbf{1}_{i \neq j} \textbf{1}_{y_i \neq y_j} log \frac{e_{ij}}{\sum_{k=1}^N \textbf{1}_{i \neq k} e_{ik}}, \notag
\end{equation}

\begin{equation}
    e_{mn} = exp(h_{scl}^m \cdot h_{scl} ^ n / \tau),
\end{equation}

The cross-entropy loss is computed by:
\begin{equation}
    L_{ce} = -\frac{1}{N} \sum_i^N (y_i log(y_i^\prime) + (1 - y_i) log (1 - y_i ^\prime)),
\end{equation}
where $N$ is the batch size, $h_{scl}$ is the embedding from the model, $N_{y_i}$ is the total number of examples with the same labels as $y_i$ in a batch and $\tau$ is the temperature parameter controlling the dissimilarity of samples from different classes should be. In cross-entropy loss, $y_i ^\prime$ is the predicted label by the model.

Finally, the loss function used to train our model is combined with $L_{scl}$ and $L_{ce}$ adjusted by a hyperparameter $\lambda$: 
\begin{equation}
    L = (1 - \lambda) L_{ce} + \lambda L_{scl} \textbf{.}
\end{equation}

\section{Experiments}

\subsection{Experimental Setup}
We perform 10-fold cross-validation and use Precision, Recall, and F1-score as metrics. They are weightedly calculated according to the number of samples in each class, which makes Recall mathematically equal to Accuracy (see Supplementary Materials for proof). Results with the best Recall among the training epochs are reported. Model implementation details are in Supplementary Materials' section B.1.

\subsection{Baselines}
We compare against 8 baselines of different specialties. Bi-LSTM~\cite{schuster1997bidirectional} is a commonly used classification model for text. TextGCN~\cite{yao2019graph}, BERT~\cite{devlin2018bert}, XLNet~\cite{yang2019xlnet} and ERNIE~\cite{sun2019ernie}, are graph-based or PLMs having shown good performance on text classification. R-GAT~\cite{wang2020relational} is a graph-based model developed for ALSC, sharing certain commonalities with our task and model in terms of task definition and information utilization. GPT-4~\cite{achiam2023gpt} and GPT-5 are also included as the baseline. EvoPrompt~\cite{guo2024connecting} is a prompt learning method that optimizes the discrete prompt via evolution algorithms to fit specific tasks. We briefly review the baselines as follows:

\begin{itemize}
    \item \textbf{Bi-LSTM} is a commonly used neural method modeling sequences. We apply a 2-layer Bi-LSTM with ReLU activation function and a linear classification layer.
    \item \textbf{TextGCN} is a GCN based model by constructing an adjacent matrix based on the co-occurrence between words and TF-IDF features, achieving better results than GCN on text based tasks. Two GCNs with ReLU are used as encoders to extract representations of life trajectory activities.
    \item \textbf{R-GAT} is a dependency-based ALSC model introduced in related work and we have it as a baseline. R-GAT uses reshape and prune strategies to focus all dependencies in the sentence on the words of interest and uses the GAT model for representation learning. We use the words in triples as the aspect word, applying the same strategies to refine the graph.
    \item \textbf{BERT} is a bidirectional encoder pre-trained language model composed of multiple transformer layers. It has been proven to be beneficial for many NLP tasks and has displayed promising results as one of the state of the art language models before LLM emerges. We use the base version of BERT and the hyperparameters recommended in~\cite{devlin2018bert} to finetune it.
    \item \textbf{XLNet} is a generalized auto-regressive pre-training method that combines BERT’s bidirectional training and Transformer-XL’s segment-level recurrence mechanism. We use the official classifier based on the base version XLNet from HuggingFace as our implementation. We use the same hyperparameters as BERT.
    \item \textbf{ERNIE} is another pre-trained language model and its improvement is from fusing entity and relation information during the pre-training process. The implementation used is also from HuggingFace. We use the base version of ERNIE and the official recommended hyperparameters to finetune it\footnote{\url{https://ai.baidu.com/ai-doc/ERNIE-Ultimate/dkhblgj0z}}. 
    \item \textbf{GPT-4} has shown outstanding performance in many language understanding tasks. We call the API of GPT-4 (gpt-4-0613) in our experiments. The prompts used for both the main results and the sensitivity study are detailed in Section~\ref{prompt_col}.
    \item \textbf{GPT-5} is OpenAI's latest integrated large language model, demonstrating superior performance over GPT-4 on complex benchmarks. The prompts used for both the main results and the sensitivity study are detailed in Section~\ref{prompt_col}.
    \item \textbf{EvoPrompt} is a prompt learning method combining the discrete prompting technique and evolution algorithm. It optimizes task specific prompts using the ability of LLM itself without the need of visiting the gradient information thus can be adopted to black box API calling. We use the Role-Playing prompt for its initialization. We use GPT-5 as the LLM backbone. 
\end{itemize}

We concatenate the triple to the end of the corresponding sentence and use a special token `[SEP]' to mark the concatenation for the PLMs.

\subsection{Experimental Results}
\textbf{Prediction Performance} As reported in Table~\ref{tab1:perf}, our model outperforms the baselines across all experiments. When transitioning from the \textit{Regular} dataset to the \textit{LLM-Refined} dataset, our method demonstrates an improvement of 2.3\% in F1 score, achieving 84.5\%, which surpasses the second-best by 1.5\%. We further recalculate the metrics by categories (9 in total, see Table~\ref{tab1}) instead of the fine-grained types (24 in total), all methods naturally reach higher F1-scores due to this relaxation, and ours hits the F1-score of 89.6\%.

Among the baselines, TextGCN, a classic graph-based text classification model, tends to underperform compared to most other methods. This may be attributed to the sparsity of the adjacency matrices constructed from word and sentence co-occurrences, as well as the oversight of triple word relationships, both of which could negatively impact its performance.

R-GAT, which is intended for the ALSC task, achieves an F1 score of only 72.2\% for type classification on \textit{LLM-Refined}, falling short of our expectations. A possible reason for this may be that R-GAT is designed to focus on a single aspect entity and its corresponding sentiment, while the triples of three entities in our scenario may exceed its capacity. 
GPT-4, GPT-5 and EvoPrompt achieve Recall under 80.0\%, suggesting prompt-based methods may struggle with this task's requirement for precise triple-context alignment. Even EvoPrompt's evolutionary prompt optimization cannot close this gap, likely because the scattered triple entities and complex syntactic relationships are difficult to capture through prompting alone.

\begin{figure*}
    \centering

    \includegraphics[width=\linewidth]{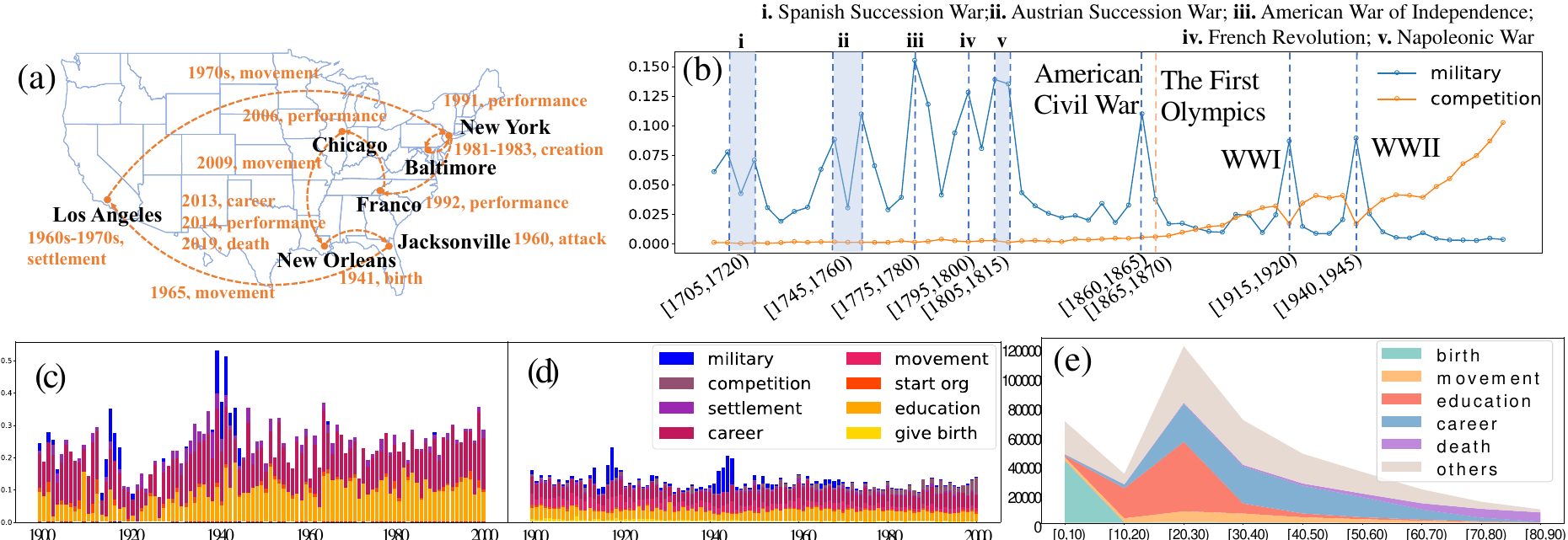}
    \caption{(a) Life trajectory of Malcolm John Rebennack Jr. (b) Ratios of military and competition activities from 1700 to 2000 in five-year intervals. Wars and the First Olympics are marked. (c) and (d) International departures from Germany and the US in the 20th century. (e) Activities in life stages in a stacked chart. X-axis marks the age group and y-axis is the number of life trajectory activities.}
    \label{fig:analysis}
\end{figure*}

\subsection{Ablation Study}

For the ablation study, we remove the syntactic MASK vector and the SCL to show their necessity. 
Meanwhile, we check the necessity of classifying the triples along with their contextual sentences, instead of classifying the sentences only, by removing the components for capturing the syntactic graphs and only keeping the ones generating sentence embedding in Figure~\ref{fig:SAM-ERNIE}. The results are shown in Table~\ref{tab:ablation_study}, as we expected, none of the ablation experiments exceed the SAM4LTC.

Compared to SAM4LTC, SAM4LTC$_{w/o\text{ }ma}$, which uses MASK that only masks the elements in the triple instead of the MASK from the local subgraph, decreases in F1 by 2.3\%. Secondly, SAM4LTC exceeds SAM4LTC$_{w/o\text{ }scl}$ that removes the supervised contrastive loss and only leaves cross-entropy loss by 2.2\% in F1. 
Finally, the SAM4LTC$_{w/o\text{ }tri}$ without triples and the relevant syntactic structures falls behind SAM4LTC by 2.1\% in F1.

\begin{table}[htbp]
\caption{Model ablation results.}
\begin{center}
\begin{tabular}{l|c@{\hspace{5pt}}c@{\hspace{5pt}}c}
    \toprule
        Model & Pre (\%) & R (Acc) (\%) & F1 (\%) \\
        \hline
        SAM4LTC$_{w/o\text{ }ma}$ & 82.2 (-3.0) & 82.9 (-1.6) & 82.2 (-2.3)\\
        SAM4LTC$_{w/o\text{ }scl}$ & 83.3 (-1.9) & 82.6 (-1.9) & 82.3 (-2.2) \\
        SAM4LTC$_{w/o\text{ }tri}$ & 83.2 (-2.0) & 82.5 (-2.0) & 82.2 (-2.1) \\
    \bottomrule
\end{tabular}
\label{tab:ablation_study}
\end{center}
\end{table}

\subsection{Sensitivity Analysis over Different Prompts}
We examine GPT-4's performance with various prompts as the classification baseline. As shown in Section D.1 of the Supplementary Materials, the F1 score ranged from 67.2\% to 73.4\%, typically below that of most baselines. We also assess how our model's performance varies with two sets of refined sentences from different GPT-4 prompts, remaining stable.

\section{Analysis of Sample Sets}
We validate our method and analyze sample sets of the labeled life trajectory activities to show the potential of the method as well as the data. We classify all trajectory triples extracted from the entire English Wikipedia, followed by necessary preprocessing (details in Supplementary Materials' Section A.3).

In all, we have a \textit{3-century} dataset of 3,828,133 triples of 589,193 people to start with. For a closer look, we narrow it down to 20th-century Germany (\textit{20th-Germany}) and the United States (\textit{20th-US}), respectively including 17,698 triples of 5,435 Germans and 229,594 triples of 75,149 Americans.

\paragraph{Data at a Glance. } From the \textit{3-century} dataset, we randomly sample 207 triples and find that 80.16\% are accurate. The type distribution of the \textit{3-century} dataset, as shown in Supplementary Materials' Section 1.3, is similar to Figure~\ref{fig:dis}. Figure~\ref{fig:analysis} (a) showcases the life trajectory of American singer. \textit{Malcolm John Rebennack Jr.} who ranks among top 20\% in the dataset, according to number of triples. Born in New Orleans, he lived in Los Angeles in the West, created his works in Baltimore and spent his main performing career in the East in the 70s and 80s. In his later years, he returned to his hometown and died in 2019. 

\paragraph{Wars and Sports in Three Centuries.} We examine two activity types, \textit{military} and \textit{competition}.
Within each 5-year range, we calculate the percentage of the specific type of activity among all activities and the overall distributions of these two activities are visualized in Figure~\ref{fig:analysis} (b). The peaks in the curve for \textit{military} correspond well with major wars temporally, from the Spanish Succession War to WWII. \textit{Competition}, on the other hand, witnesses a relatively steady growth, starting from the late 19th century, consistent with the expansion of popular sports, marked by the first modern Olympics in 1896. Curiously, the \textit{competition} trend obviously drops during WWI and WWII, and it actually negatively correlates with the \textit{military} trend during the 20th century (Pearson correlation coefficient being $-0.52$, with a p-value of $2.42e{-5}$).

\paragraph{How/Why People Leave Home Countries.} We zoom in and conduct fine-grained analysis on Germany and the United States to examine how and why people depart their home countries. We select the top 8 types except birth and death for the analysis. An international departure is a move from the home country to a foreign one, judged by two consecutive trajectory triples of a same person. Figure~\ref{fig:analysis} (c) and (d) visualize the ratios of international departures against all travels started within the home country (including international departures and the travels within the home country), for the two countries respectively.

For Germany, we see a surge in international departures around the 1930s and early 1940s which may be associated with the rule of Nazi. If we ignore the \textit{military} activities in blue, the surge still persists. Its larger contributions are from the increase in \textit{career} and \textit{settlement}, which may indicate a brain drain. 
In fact, during this period, many Jewish figures, such as Albert Einstein, Hannah Arendt (philosopher), and Kurt Weill (composer) fled Germany. 
If we look at the US, its ratio of international departures is relatively lower and more stable. Its two surges correspond to the two world wars. Interestingly, \textit{military} activities mainly contribute to these two peaks and the US does not seem to suffer from brain drain.

\paragraph{Individual Life Stages and Activities.} We further focus on the Americans in the 20th century and examine their life stages and activities aggregated at the individual level. We visualize different trajectory activities throughout different life stages, with a stacked chart aggregating these people's activities (Figure~\ref{fig:analysis} (e)). In each age group, we show the total amounts for the \textit{birth}, \textit{death}, \textit{education}, \textit{career}, and \textit{movement} types of activities, which account for about 70\% samples of the dataset. Rest activities are labeled as \textit{others}. It shows that people in their 20s have the most trajectory activities, with the top ones being \textit{education} and \textit{career}, while the death type takes up larger portions beyond age 70.

\paragraph{Movement Dynamics of Different Activity Types} 
We additionally analyze the life trajectory activities of \textit{education} and \textit{career}, referencing perspectives from the milestone work of cultural history analysis~\cite{schich2014network}. This narrows down the samples to 25,031 individuals with both activities.

\begin{figure}
    \centering
    \includegraphics[width=\linewidth]{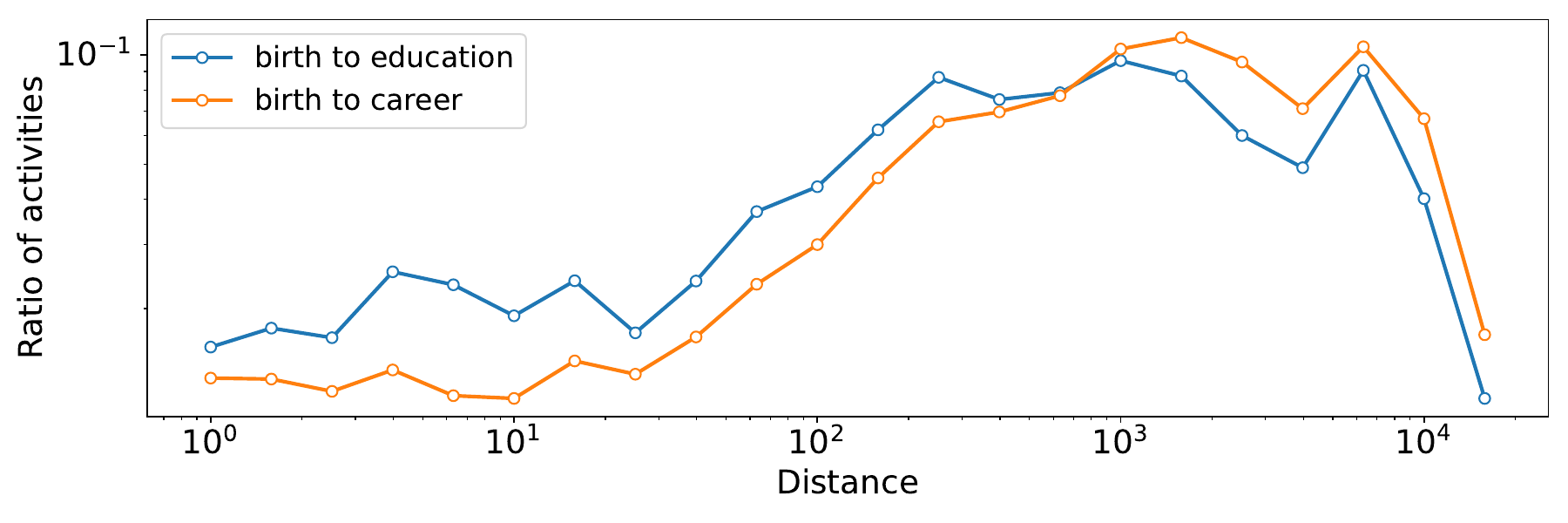}
    \caption{Distribution of distance from locations of birth to education/career.}
    \label{fig:distance_distribution}
\end{figure}

We compute the distance between the education place and the birthplace of the corresponding person for each \textit{education} activity. In Figure~\ref{fig:distance_distribution}, we plot the distribution of birth-to-education distances (in blue) and do the same for career activities (in orange). We observe that more people choose to receive education within hundreds of kilometers from their birthplaces, while more people go further to pursue their career and the overall average distance to education is much shorter than that to work $(1866km < 2541km)$ in the 20th-century US. 

\section{Conclusion}
We propose a new task of labeling life trajectory activities of people and introduce the taxonomy including 24 types in 9 categories of life trajectory activities with a manually annotated dataset. We design SAM4LTC fusing syntactic structure into text representation to classify life trajectory activities and use LLM to refine the syntactic structures of sentences. Our model outperforms the introduced baselines in the experiments. Besides, we conduct an empirical analysis using the life trajectory activities with corresponding types. We hope our code and the largest fine-grained life trajectory dataset will facilitate studies on the grand narratives of human dynamics, both analytically and interactively, while also offering insights into information extraction.

\clearpage
\newpage
\bibliographystyle{named}
\bibliography{ijcai26}

@inproceedings{fei2022inheriting,
  title={Inheriting the Wisdom of Predecessors: A Multiplex Cascade Framework for Unified Aspect-based Sentiment Analysis.},
  author={Fei, Hao and Li, Fei and Li, Chenliang and Wu, Shengqiong and Li, Jingye and Ji, Donghong},
  booktitle={IJCAI},
  pages={4121--4128},
  year={2022}
}

@article{zhang2024paths,
  title={Paths of A Million People: Extracting Life Trajectories from Wikipedia},
  author={Zhang, Ying and Li, Xiaofeng and Liu, Zhaoyang and Zhang, Haipeng},
  journal={arXiv preprint arXiv:2406.00032},
  year={2024}
}

@inproceedings{doddington2004automatic,
  title={The automatic content extraction (ace) program-tasks, data, and evaluation.},
  author={Doddington, George R and Mitchell, Alexis and Przybocki, Mark A and Ramshaw, Lance A and Strassel, Stephanie M and Weischedel, Ralph M},
  booktitle={LREC},
  volume={2},
  pages={837--840},
  year={2004},
  organization={Lisbon}
}

@article{schich2014network,
  title={A network framework of cultural history},
  author={Schich, Maximilian and Song, Chaoming and Ahn, Yong-Yeol and Mirsky, Alexander and Martino, Mauro and Barab{\'a}si, Albert-L{\'a}szl{\'o} and Helbing, Dirk},
  journal={Science},
  volume={345},
  number={6196},
  pages={558--562},
  year={2014},
  publisher={American Association for the Advancement of Science}
}

@article{creativityovertime,
  title={Creativity over time and space: A historical analysis of European cities},
  author={Serafinelli, Michel and Tabellini, Guido},
  journal={Journal of Economic Growth},
  volume={27},
  number={1},
  pages={1--43},
  year={2022},
  publisher={Springer}
}

@article{elder1994time,
  title={Time, human agency, and social change: Perspectives on the life course},
  author={Elder Jr, Glen H},
  journal={Social psychology quarterly},
  pages={4--15},
  year={1994},
  publisher={JSTOR}
}

@article{wang2020relational,
  title={Relational graph attention network for aspect-based sentiment analysis},
  author={Wang, Kai and Shen, Weizhou and Yang, Yunyi and Quan, Xiaojun and Wang, Rui},
  journal={arXiv preprint arXiv:2004.12362},
  year={2020}
}

@incollection{schatzki2022trajectories,
  author={{Schatzki, Theodore R}},
  title={The trajectories of a life},
  booktitle={Doing Transitions in the Life Course: Processes and Practices},
  pages={19--34},
  year={2022},
  publisher={Springer International Publishing Cham}
}

@book{schatzki2019social,
  title={Social change in a material world},
  author={Schatzki, Theodore R},
  year={2019},
  publisher={Routledge}
}

@article{verginer2020cities,
  title={Cities and countries in the global scientist mobility network},
  author={Verginer, Luca and Riccaboni, Massimo},
  journal={Applied Network Science},
  volume={5},
  pages={1--16},
  year={2020},
  publisher={Springer}
}

@article{janssen2011age,
  title={Age effects in cultural life scripts},
  author={Janssen, Steve MJ and Rubin, David C},
  journal={Applied Cognitive Psychology},
  volume={25},
  number={2},
  pages={291--298},
  year={2011},
  publisher={Wiley Online Library}
}

@article{khosla2020supervised,
  title={Supervised contrastive learning},
  author={Khosla, Prannay and Teterwak, Piotr and Wang, Chen and Sarna, Aaron and Tian, Yonglong and Isola, Phillip and Maschinot, Aaron and Liu, Ce and Krishnan, Dilip},
  journal={NeurIPS},
  volume={33},
  pages={18661--18673},
  year={2020}
}

@article{zhang2018generalized,
  title={Generalized cross entropy loss for training deep neural networks with noisy labels},
  author={Zhang, Zhilu and Sabuncu, Mert},
  journal={NeurIPS},
  volume={31},
  year={2018}
}

@inproceedings{dickinson2015identifying,
  title={Identifying prominent life events on twitter},
  author={Dickinson, Thomas and Fernandez, Miriam and Thomas, Lisa A and Mulholland, Paul and Briggs, Pam and Alani, Harith},
  booktitle={K-CAP},
  pages={1--8},
  year={2015}
}

@inproceedings{yen2019personal-bak,
  title={Personal knowledge base construction from text-based lifelogs},
  author={Yen, An-Zi and Huang, Hen-Hsen and Chen, Hsin-Hsi},
  booktitle={SIGIR},
  pages={185--194},
  year={2019}
}

@article{devlin2018bert,
  title={Bert: Pre-training of deep bidirectional transformers for language understanding},
  author={Devlin, Jacob and Chang, Ming-Wei and Lee, Kenton and Toutanova, Kristina},
  journal={arXiv preprint arXiv:1810.04805},
  year={2018}
}

@article{herrera2023ballet,
  title={Quantifying hierarchy and prestige in US ballet academies as social predictors of career success},
  author={Herrera-Guzm{\'a}n, Yessica and Gates, Alexander J and Candia, Cristian and Barab{\'a}si, Albert-L{\'a}szl{\'o}},
  journal={Scientific Reports},
  volume={13},
  number={1},
  pages={18594},
  year={2023},
  publisher={Nature Publishing Group UK London}
}

@article{huang2020careerhistorical,
  title={Historical comparison of gender inequality in scientific careers across countries and disciplines},
  author={Huang, Junming and Gates, Alexander J and Sinatra, Roberta and Barab{\'a}si, Albert-L{\'a}szl{\'o}},
  journal={PNAS},
  volume={117},
  number={9},
  pages={4609--4616},
  year={2020},
  publisher={National Acad Sciences}
}

@article{ke2021incorporating,
  title={Incorporating explicit syntactic dependency for aspect level sentiment classification},
  author={Ke, Wenjun and Gao, Jinhua and Shen, Huawei and Cheng, Xueqi},
  journal={Neurocomputing},
  volume={456},
  pages={394--406},
  year={2021},
  publisher={Elsevier}
}

@inproceedings{huang2020syntax,
  title={Syntax-aware graph attention network for aspect-level sentiment classification},
  author={Huang, Lianzhe and Sun, Xin and Li, Sujian and Zhang, Linhao and Wang, Houfeng},
  booktitle={COLING},
  pages={799--810},
  year={2020}
}

@inproceedings{liu2021can,
  title={Can language models identify Wikipedia articles with readability and style issues?},
  author={Liu, Yang and Medlar, Alan and Glowacka, Dorota},
  booktitle={SIGIR},
  pages={113--117},
  year={2021}
}

@inproceedings{lai2020event,
  title={Event Detection: Gate Diversity and Syntactic Importance Scores for Graph Convolution Neural Networks},
  author={Lai, Viet Dac and Nguyen, Tuan Ngo and Nguyen, Thien Huu},
  booktitle={EMNLP},
  pages={5405--5411},
  year={2020}
}

@article{park2019global,
  title={Global labor flow network reveals the hierarchical organization and dynamics of geo-industrial clusters},
  author={Park, Jaehyuk and Wood, Ian B and Jing, Elise and Nematzadeh, Azadeh and Ghosh, Souvik and Conover, Michael D and Ahn, Yong-Yeol},
  journal={Nature communications},
  volume={10},
  number={1},
  pages={3449},
  year={2019},
  publisher={Nature Publishing Group UK London}
}

@article{schouten2015survey,
  title={Survey on aspect-level sentiment analysis},
  author={Schouten, Kim and Frasincar, Flavius},
  journal={TKDE},
  volume={28},
  number={3},
  pages={813--830},
  year={2015},
  publisher={IEEE}
}

@article{schuster1997bidirectional,
  title={Bidirectional recurrent neural networks},
  author={Schuster, Mike and Paliwal, Kuldip K},
  journal={IEEE Transactions on Signal Processing},
  volume={45},
  number={11},
  pages={2673--2681},
  year={1997},
  publisher={Ieee}
}

@inproceedings{yao2019graph,
  title={Graph convolutional networks for text classification},
  author={Yao, Liang and Mao, Chengsheng and Luo, Yuan},
  booktitle={AAAI},
  volume={33},
  pages={7370--7377},
  year={2019}
}

@article{yang2019xlnet,
  title={Xlnet: Generalized autoregressive pretraining for language understanding},
  author={Yang, Zhilin and Dai, Zihang and Yang, Yiming and Carbonell, Jaime and Salakhutdinov, Russ R and Le, Quoc V},
  journal={NeurIPS},
  volume={32},
  year={2019}
}

@article{achiam2023gpt,
  title={Gpt-4 technical report},
  author={Achiam, Josh and Adler, Steven and Agarwal, Sandhini and Ahmad, Lama and Akkaya, Ilge and Aleman, Florencia Leoni and Almeida, Diogo and Altenschmidt, Janko and Altman, Sam and Anadkat, Shyamal and others},
  journal={arXiv preprint arXiv:2303.08774},
  year={2023}
}

@inproceedings{moller2024parrot,
  title={The parrot dilemma: Human-labeled vs. llm-augmented data in classification tasks},
  author={M{\o}ller, Anders Giovanni and Pera, Arianna and Dalsgaard, Jacob and Aiello, Luca},
  booktitle={EACL},
  pages={179--192},
  year={2024}
}

@inproceedings{whitehouse2023llm,
  title={LLM-powered Data Augmentation for Enhanced Cross-lingual Performance},
  author={Whitehouse, Chenxi and Choudhury, Monojit and Aji, Alham Fikri},
  booktitle={EMNLP},
  year={2023}
}

@article{atari2025chronospatial,
  title={The chronospatial revolution in psychology},
  author={Atari, Mohammad and Henrich, Joseph and Schulz, Jonathan},
  journal={Nature Human Behaviour},
  pages={1--9},
  year={2025},
  publisher={Nature Publishing Group UK London}
}

@article{li2023synthetic,
  title={Synthetic data generation with large language models for text classification: Potential and limitations},
  author={Li, Zhuoyan and Zhu, Hangxiao and Lu, Zhuoran and Yin, Ming},
  journal={arXiv preprint arXiv:2310.07849},
  year={2023}
}

@misc{xu2025largereasoningmodelssurvey,
      title={Towards Large Reasoning Models: A Survey of Reinforced Reasoning with Large Language Models}, 
      author={Fengli Xu and Qianyue Hao and Zefang Zong and Jingwei Wang and Yunke Zhang and Jingyi Wang and Xiaochong Lan and Jiahui Gong and Tianjian Ouyang and Fanjin Meng and Chenyang Shao and Yuwei Yan and Qinglong Yang and Yiwen Song and Sijian Ren and Xinyuan Hu and Yu Li and Jie Feng and Chen Gao and Yong Li},
      year={2025},
      eprint={2501.09686},
      archivePrefix={arXiv},
      primaryClass={cs.AI},
      url={https://arxiv.org/abs/2501.09686}, 
}

@article{vaswani2017attention,
  title={Attention is all you need},
  author={Vaswani, A},
  journal={Advances in Neural Information Processing Systems},
  year={2017}
}

@article{aleksanyan2021state,
  title={Do state visits affect cross-border mergers and acquisitions?},
  author={Aleksanyan, Mark and Hao, Zhiwei and Vagenas-Nanos, Evangelos and Verwijmeren, Patrick},
  journal={Journal of Corporate Finance},
  volume={66},
  pages={101800},
  year={2021},
  publisher={Elsevier}
}

@article{wang2019early,
  title={Early-career setback and future career impact},
  author={Wang, Yang and Jones, Benjamin F and Wang, Dashun},
  journal={Nature communications},
  volume={10},
  number={1},
  pages={4331},
  year={2019},
  publisher={Nature Publishing Group UK London}
}

@article{wolfinger2008problems,
  title={Problems in the pipeline: Gender, marriage, and fertility in the ivory tower},
  author={Wolfinger, Nicholas H and Mason, Mary Ann and Goulden, Marc},
  journal={The Journal of Higher Education},
  volume={79},
  number={4},
  pages={388--405},
  year={2008},
  publisher={Taylor \& Francis}
}

@article{quarles2020shape,
  title={The shape of educational inequality},
  author={Quarles, Christopher L and Budak, Ceren and Resnick, Paul},
  journal={Science Advances},
  volume={6},
  number={29},
  pages={eaaz5954},
  year={2020},
  publisher={American Association for the Advancement of Science}
}

@article{mukherjee2019prior,
  title={Prior shared success predicts victory in team competitions},
  author={Mukherjee, Satyam and Huang, Yun and Neidhardt, Julia and Uzzi, Brian and Contractor, Noshir},
  journal={Nature human behaviour},
  volume={3},
  number={1},
  pages={74--81},
  year={2019},
  publisher={Nature Publishing Group UK London}
}

@article{minaee2021deep,
  title={Deep learning--based text classification: a comprehensive review},
  author={Minaee, Shervin and Kalchbrenner, Nal and Cambria, Erik and Nikzad, Narjes and Chenaghlu, Meysam and Gao, Jianfeng},
  journal={ACM computing surveys (CSUR)},
  volume={54},
  number={3},
  pages={1--40},
  year={2021},
  publisher={ACM New York, NY, USA}
}

@article{viberg1983verbs,
  title={The verbs of perception: A typological study},
  author={Viberg, {\AA}ke},
  journal={Explanations for Language Universals},
  pages={123},
  year={1983},
  publisher={Walter de Gruyter, Berlin/New York Berlin, New York}
}

@article{sun2019ernie,
  title={Ernie: Enhanced representation through knowledge integration},
  author={Sun, Yu and Wang, Shuohuan and Li, Yukun and Feng, Shikun and Chen, Xuyi and Zhang, Han and Tian, Xin and Zhu, Danxiang and Tian, Hao and Wu, Hua},
  journal={arXiv preprint arXiv:1904.09223},
  year={2019}
}

@article{wang2018glue,
  title={GLUE: A multi-task benchmark and analysis platform for natural language understanding},
  author={Wang, Alex and Singh, Amanpreet and Michael, Julian and Hill, Felix and Levy, Omer and Bowman, Samuel R},
  journal={arXiv preprint arXiv:1804.07461},
  year={2018}
}

@article{ho2022large,
  title={Large language models are reasoning teachers},
  author={Ho, Namgyu and Schmid, Laura and Yun, Se-Young},
  journal={arXiv preprint arXiv:2212.10071},
  year={2022}
}

@inproceedings{zhang2020convolution,
  title={Convolution over hierarchical syntactic and lexical graphs for aspect level sentiment analysis},
  author={Zhang, Mi and Qian, Tieyun},
  booktitle={EMNLP},
  pages={3540--3549},
  year={2020}
}

@article{dai2021does,
  title={Does syntax matter? a strong baseline for aspect-based sentiment analysis with roberta},
  author={Dai, Junqi and Yan, Hang and Sun, Tianxiang and Liu, Pengfei and Qiu, Xipeng},
  journal={arXiv preprint arXiv:2104.04986},
  year={2021}
}

@article{zhang2021model,
  title={Model-based clustering of time-dependent categorical sequences with application to the analysis of major life event patterns},
  author={Zhang, Yingying and Melnykov, Volodymyr and Zhu, Xuwen},
  journal={Statistical Analysis and Data Mining: The ASA Data Science Journal},
  volume={14},
  number={3},
  pages={230--240},
  year={2021},
  publisher={Wiley Online Library}
}

@inproceedings{qian2023open,
  title={Open-world social event classification},
  author={Qian, Shengsheng and Chen, Hong and Xue, Dizhan and Fang, Quan and Xu, Changsheng},
  booktitle={Proceedings of the ACM Web Conference 2023},
  pages={1562--1571},
  year={2023}
}

@inproceedings{park2023generative,
  title={Generative agents: Interactive simulacra of human behavior},
  author={Park, Joon Sung and O'Brien, Joseph and Cai, Carrie Jun and Morris, Meredith Ringel and Liang, Percy and Bernstein, Michael S},
  booktitle={Proceedings of the 36th annual acm symposium on user interface software and technology},
  pages={1--22},
  year={2023}
}

@misc{bai2024baijia,
      title={BaiJia: A Large-Scale Role-Playing Agent Corpus of Chinese Historical Characters}, 
      author={Ting Bai and Jiazheng Kang and Jiayang Fan},
      year={2025},
      eprint={2412.20024},
      archivePrefix={arXiv},
      primaryClass={cs.AI},
      url={https://arxiv.org/abs/2412.20024}, 
}

@article{shanahan2023role,
  title={Role play with large language models},
  author={Shanahan, Murray and McDonell, Kyle and Reynolds, Laria},
  journal={Nature},
  volume={623},
  number={7987},
  pages={493--498},
  year={2023},
  publisher={Nature Publishing Group UK London}
}

@inproceedings{
guo2024connecting,
title={Connecting Large Language Models with Evolutionary Algorithms Yields Powerful Prompt Optimizers},
author={Qingyan Guo and Rui Wang and Junliang Guo and Bei Li and Kaitao Song and Xu Tan and Guoqing Liu and Jiang Bian and Yujiu Yang},
booktitle={The Twelfth International Conference on Learning Representations},
year={2024},
url={https://openreview.net/forum?id=ZG3RaNIsO8}
}

\newpage
\appendix

\section{Dataset}

\subsection{Further Descriptions of the Types of Life Trajectory Activities}

We will further supplement the explanation of the taxonomy of life trajectory activities here.

\begin{enumerate}

    \item Life
    \begin{itemize}[label={}]
        \item (1) Birth: Someone's life starts.
        \item (2) Death: Someone's life ends for any reason.
        \item (3) Marriage: Two people get married or engage.
        \item (4) Injure and Illness: Affairs related injures and illnesses such as receiving treatments.
        \item (5) Settlement: Someone resides in a certain place for a long or short period of time.
        \item (6) Education: Education events such as admission, enrollment, or graduation.
        \item (7) Give birth: A mother gives birth to a baby.
        \item (8) Accident: Unexpected events causing losses such as car accident, falling from a height.
        \item (9) Purchase and Sell: Generally involves purchasing and selling large assets such as real estate.
        \item (10) Divorce: Two people get divorced.
    \end{itemize}
    
    \item Career
    \begin{itemize}[label={}]
        \item (11) Career: Event related to career changes such as being promoted or demoted, start or end a position.
    \end{itemize}

    \item Pro-Event (Professional Event)
    \begin{itemize}[label={}]
        \item (12) Competition: Someone participates in a competition such as winning or losing a competition.
        \item (13) Performance: Someone plays a role in a performance, such as playing a role in a play, conducting or playing instruments in a concert.
        \item (14) Exhibition: Organizing personal exhibitions.
        \item (15) Creation: Do creative jobs, such as writing a book, recording an album and drawing a painting.
        \item (16) Campaign: Activities or efforts to hold a certain position such as running for president of a state or an institute.
        \item (17) Start org: Start an organization such as starting a bar and opening a company.
    \end{itemize}

    \item Contact
    \begin{itemize}[label={}]
        \item (18) Meet: Small scale gatherings such as receiving interviews and encounters.
        \item (19) Assembly: Large scale gatherings for certain purposes such as attending a conference or award ceremony, giving a speech and protesting.
    \end{itemize}
    
    \item Justice
    \begin{itemize}[label={}]
        \item (20) Justice: Everything related to justice such as arrests, trials, parole and testimony.
    \end{itemize}

    \item Attack 
    \begin{itemize}[label={}]
        \item (21) Attack: The act of destroying, damaging, or harming a target such as shooting, street fights and raids.
    \end{itemize}
    
    \item Movement
    \begin{itemize}[label={}]
        \item (22) Attack: Someone moves from one place to another.
    \end{itemize}
    
    \item Military
    \begin{itemize}[label={}]
        \item (23) Military: Life trajectory activities related to military events, such as an military action or serving in the army.
    \end{itemize}
    
    \item Other
    \begin{itemize}[label={}]
        \item (24) Other: Life trajectory activities that are hard to be applied to any types above and almost meaningless such as staying in a certain place without further details.
    \end{itemize}

\end{enumerate}

As mentioned in the Introduction, a context sentence may contain multiple types of life trajectories, but only one of them is associated with the extracted trajectory activity triple. Therefore, during the annotation process, we annotate each triple according to the type of life trajectory associated with the triple ignoring the rest types.

\subsection{Statistics of the \textit{Regular} Dataset and the \textit{3-century} dataset}

\begin{figure}
    \centering
    \includegraphics[width=\linewidth]{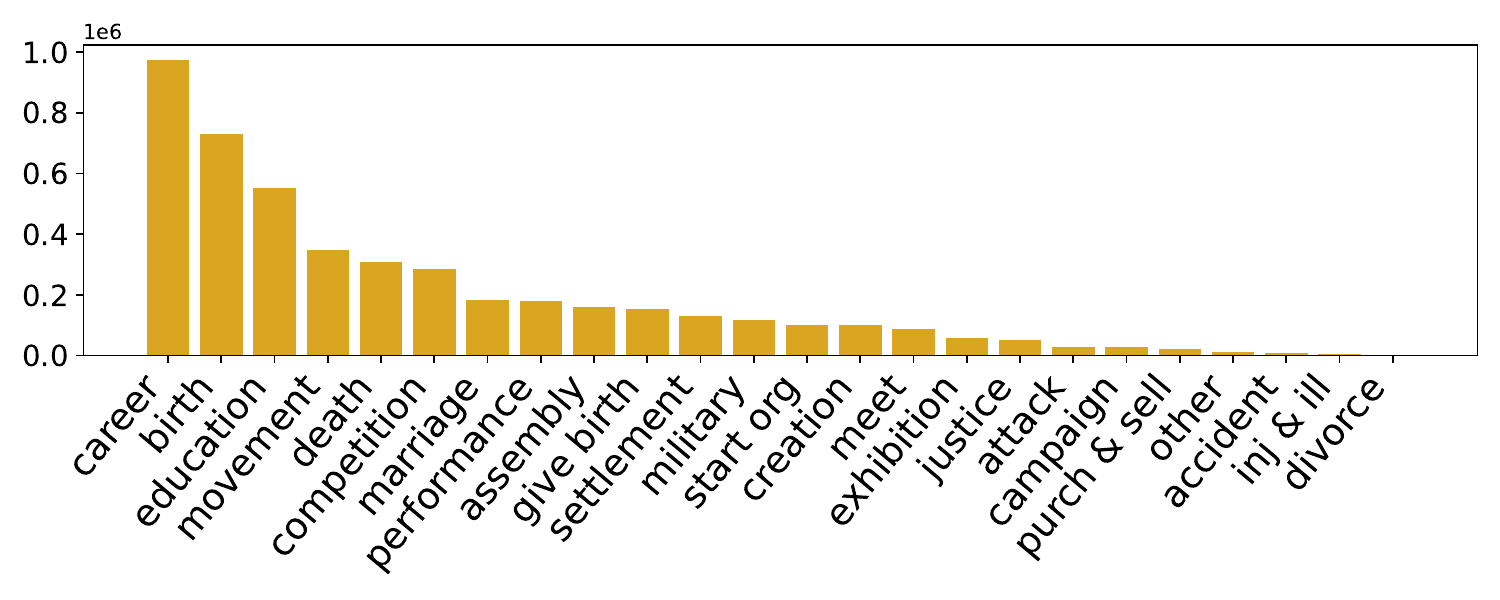}
    \caption{Type distribution in the \textit{3-century} dataset.}
    \label{fig:dis_all}
\end{figure}

The distributions of the manually annotated \textit{Regular} dataset and the \textit{3-century} dataset are shown in main text's Figure 2 and Figure~\ref{fig:dis_all} here, respectively. The type distributions of the two datasets are consistent. It can be seen that the most frequent life trajectory activity type is \textit{career}, followed by \textit{birth}, \textit{education}, \textit{movement} and \textit{death}.

\subsection{Data Selection and Preprocessing of \textit{3-century} dataset}
We initially extract all life trajectories within three centuries, from 1700 to 2000. We then keep people with more than 3 extracted life trajectory triples, and use SAM4LTC to label the type of each life trajectory triple. To improve data quality, we use a co-reference tool\footnote{\url{https://huggingface.co/models?other=coreference-resolution}} on HuggingFace to replace the third-person pronoun (`he' or `she') in a triple with the person name that it refers to in the context. Additionally, we use Nominatim\footnote{\url{https://nominatim.org/}} to do the geocoding.

\section{Experiments}

\subsection{Model Implementations} We use the base version of ERNIE whose output embedding is $d = 768$ and set $\lambda = 0.7$ and $\tau = 0.1$ in loss terms. We train SAM4LTC using an AdamW optimizer with configuration of $1e^{-5}$ learning rate. All experiments above are conducted on two RTX 3090. The refinement is implemented by calling the API of GPT-4 (gpt-4-0613).

\subsection{Simple Deviation of the equality of Weighted Recall and Accuracy}

Let the $TP_i$ be the number of true positive sample of the $i$-th class in the confusion matrix, and $FN_i$, $TN_i$ and $FP_i$ be the false negative, true negative and false positive. 

The accuracy (Acc) is  calculated as:

\begin{align}
\text{Acc} = \frac{\sum_{i=1}^{n} TP_i}{\sum_{i=1}^{n} (TP_i + FP_i + FN_i + TN_i)}
\end{align}

The weighted recall (WR) is calculated as:

\begin{align}
    \text{WR} &= \sum_{i=1}^{n} W_i \times \text{Recall}_i \\ \nonumber
    &= \sum_{i=1}^{n} \left( \frac{S_i}{\sum_{j=1}^{n} S_j} \times \frac{TP_i}{TP_i + FN_i} \right) 
\end{align}
where $S_i$ is the number of samples in $i$-th class in the test set and $\text{Recall}_i$ is the recall of the $i$-th class.

If we use $TP_i$, $FN_i$, $TN_i$ and $FP_i$ to represent $S_i$:

\begin{align}
    \text{WR} &= \sum_{i=1}^{n} \left( \frac{TP_i + FN_i}{\sum_{j=1}^{n} (TP_j + FN_j)} \times \frac{TP_i}{TP_i + FN_i} \right) \\ \nonumber
    &=\sum_{i=1}^{n} \frac{TP_i}{\sum_{j=1}^{n} (TP_j + FN_j)}
\end{align}

Therefore, we have
\begin{align}
    \text{WR} = \frac{\sum_{i=1}^{n} TP_i}{N} = \text{Acc}
\end{align}
where $N = \sum_{i=1}^{n} (TP_i + FP_i + FN_i + TN_i)$.

\section{Sensitivity Analysis over Different Prompts}

We test performance changes when using different prompts. First, we examine the performance variations of GPT-4 as the classification baseline. Additionally, we assess how our model's performance shifts with various refined sentences as input. The prompts are designed according to the tasks and adhere to several prompt engineering guidelines\footnote{\url{https://www.promptingguide.ai/}}.

\subsection{Sensitivity Analysis of GPT-4 Baseline}

As listed in Section~\ref{prompt_col}, we curate three sets of prompts for this classification task. The first is the Few-Shot prompt, which is one of the most commonly used prompting methods for LLMs and is also the one employed for the results in the main text. The second is the Multiple-Choice style prompt, aligning with the definition of multi-classification. The third is the role-playing prompt. We choose not to use the popular Chain-of-Thought prompt, as the task does not involve clear step-by-step logical reasoning; rather, it requires aggregating scattered information to determine the type for the triple.

Table~\ref{tab:GPT4sensiticity} shows that, across the three sets of prompts, GPT-4’s F1 score ranged from 67.2\% to 73.4\%. Overall, GPT-4's performance is generally below that of most baselines.
\begin{table}[htbp]
\caption{Performance of the GPT-4 baseline over different prompts.}
\begin{center}
\begin{tabular}{l|c@{\hspace{5pt}}c@{\hspace{5pt}}c}
    \toprule
        Prompt & Pre & R (Acc) & F1  \\
        \hline
        Few-Shot (main results) & 74.7\% & 74.6\% & 72.9\%\\
        Multiple Choice & 78.3\% & 75.5\% & 73.4\% \\
        Role-Playing & 73.4\% & 70.8\% & 67.2\% \\ 
    \bottomrule
\end{tabular}
\label{tab:GPT4sensiticity}
\end{center}
\end{table}

\subsection{Sensitivity Analysis of GPT-4 Refinement}

We assess our model's performance on various refined sentences as input. We curate prompts in two different styles: the naive Requirement List style (used for the main results) and the Role-Playing style. The prompt details can be found in Section~\ref{prompt_col}. We also do not use the Chain-of-Thought prompt due to the nature of the task, which does not involve clear step-by-step logical reasoning. As shown in Table~\ref{tab:GPT4sensiticityrefine}, our model demonstrates similar performance across both input sets.

\begin{table}[htbp]
\caption{Performance of our model on refined sentences generated by GPT-4 using different prompts.}
\begin{center}
\begin{tabular}{l|c@{\hspace{5pt}}c@{\hspace{5pt}}c}
    \toprule
        Prompt & Pre & R (Acc) & F1  \\
        \hline
        Req List (main results) & {84.6\%} & {85.4\%} & {84.4\%}\\
        Role-Playing & 85.8\% & 85.0\% & 84.8\% \\
    \bottomrule
\end{tabular}
\label{tab:GPT4sensiticityrefine}
\end{center}
\end{table}

\subsection{Prompts}\label{prompt_col}

\subsubsection{Prompts for GPT-4 and GPT-5 Baseline}
\paragraph{Few-Shot}

Here are a sentence and a triple.

    Sentence: \{sentence\}
    
    Triple: \{triple\}

    The triple indicates that who at what time is in somewhere described in the sentence.
    
    Based on the triple and the sentence, which class does the sentence belong to?
    
    Here are the classes you should choose from:
    \{label space\}

    For example, you should output `Career' when given the sentence and triple:
    
    Sentence: From 1946 to 1948 he was flute professor at Kneller Hall.
    
    Triple: (he, From 1946 to 1948, Kneller Hall)

    You should output `Military' when given the sentence and triple:
    
    Sentence: In 1961, Hoare's first mercenary action was in Katanga, a province trying to rebel from the newly independent Republic of the Congo.
    
    Triple: (Hoare, 1961, Katanga)

    You only need to return the class without any additional output.

\paragraph{Multiple Choice}
    Sentence: \{sentence\}
    
    Question: What does \{person\} do at \{time\} in \{location\}?

    Choose from the list:
        \{label space\}

    For example, you should choose `Career' when given
    
    Sentence: From 1946 to 1948 he was flute professor at Kneller Hall.
    Triple: (he, From 1946 to 1948, Kneller Hall)

    You should choose `Military' when given
    
    Sentence: In 1961, Hoare's first mercenary action was in Katanga, a province trying to rebel from the newly independent Republic of the Congo.
    
    Triple: (Hoare, 1961, Katanga)

    You only need to return the word in the list without any additional output.

\paragraph{Role-Playing}
You are a detective. now you are given a triple of (person, time, location) and a sentence containing the triple.

    Choose a proper activity type from the type list \{label space\} and return me the type.

    Sentence: \{sentence\}
    
    Triple: \{triple\}

    For example, you should choose `Career' when given
    
    Sentence: From 1946 to 1948 he was flute professor at Kneller Hall.
    
    Triple: (he, From 1946 to 1948, Kneller Hall)

    You should choose `Military' when given
    
    Sentence: In 1961, Hoare's first mercenary action was in Katanga, a province trying to rebel from the newly independent Republic of the Congo.
    
    Triple: (Hoare, 1961, Katanga)

    You only need to return the type from the type list without any additional output.

\subsubsection{Prompts for Refinement}

\paragraph{Requirement List}
Please rewrite the sentence according to the requirements.

    Requirements:
    
     1. Do not change the original meaning of the sentence
     
     2. Use the exactly same words in [\{person\}, \{time\} \{location\}]
     
     3. You can add, reduce, or modify some other words or phrases, such as verbs
     
     4. The rewritten sentence cannot be the same as the original sentence
     
    Sentence that needs to be rewritten: \{sentence\}.

    You may remove some information unrelated to the given person, time and location.
    
    You only need to return me the rewritten sentence without any additional output.
    
\paragraph{Role-playing}

You are an editor, You will be provided a sentence describing `person' do something in `time' in `location'.

    Please rewrite the sentence: \{sentence\}.
    
    Highlight what \{person\} (person) do in \{time\} (time) and \{location\} (location).

    You MUST use the exact word of
    
    person: \{person\};
    
    time: \{time\};
    
    location: \{location\}.

    You may remove some information unrelated to the given person, time and location
    
    You only need to return me the refined sentence without any additional output.

\end{document}